# Does Banque de France control inflation and unemployment?

Ivan O. Kitov
Russian Academy of Sciences

Oleg I. Kitov
The University of Oxford

**Abstract**
We re-estimate statistical properties and predictive power of a set of Phillips curves, which are expressed as linear and lagged relationships between the rates of inflation, unemployment, and change in labour force. For France, several relationships were estimated eight years ago. The change rate of labour force was used as a driving force of inflation and unemployment within the Phillips curve framework. Following the original problem formulation by Fisher and Phillips, the set of nested models starts with a simplistic version without autoregressive terms and one lagged term of explanatory variable. The lag is determined empirically together with all coefficients. The model is estimated using the Boundary Element Method (BEM) with the least squares method applied to the integral solutions of the differential equations. All models include one structural break might be associated with revisions to definitions and measurement procedures in the 1980s and 1990s as well as with the change in monetary policy in 1994-1995. For the GDP deflator, our original model provided a root mean squared forecast error (RMSFE) of 1.0% per year at a four-year horizon for the period between 1971 and 2004. The same RMSFE is estimated with eight new readings obtained since 2004. The rate of CPI inflation is predicted with RMSFE=1.5% per year. For the naive (no change) forecast, RMSFE at the same time horizon is 2.95% and 3.3% per year, respectively. Our model outperforms the naive one by a factor of 2 to 3. The relationships for inflation were successfully tested for cointegration. We have formally estimated several vector error correction (VEC) models for two measures of inflation. In the VAR representation, these VECMs are similar to the Phillips curves. At a four year horizon, the estimated VECMs provide significant statistical improvements on the results obtained by the BEM: RMSFE=0.8% per year for the GDP deflator and ~1.2% per year for CPI. For a two year horizon, the VECMs improve RMSFEs by a factor of 2, with the smallest RMSFE=0.5% per year for the GDP deflator. This study has validated the reliability and accuracy of the linear and lagged relationships between inflation, unemployment, and the change in labour force between 1970 and 2012.

## 1. Introduction

Price stability was chosen as a single objective by the European System of Central Banks (ECB, 2004). In quantitative terms, the Banque de France has to conduct monetary policy retaining the year-on-year increase in the Harmonized Index of Consumer Prices (HICP) below 2% (Banque de France, 2004). To achieve price stability, monetary policy should be founded on a solid ground of theoretical and empirical knowledge of the driving forces behind price inflation. In this study, we revise our model of inflation in France (Kitov, 2007), which was based on the concept of inflation developed for the USA (Kitov, 2006). Our approach stems from the original Phillips curve (1958) and from the results obtained by Fisher (1926). The only macroeconomic variable we tested as a predictor of inflation and unemployment was the change in the level of workforce. Our modelling spans the period from 1970 to 2012 and includes one structural break, which is estimated together with other defining coefficients.

Two major aspects of the broader inflation research are addressed: the cause of price change and forecasting accuracy. Firstly, we prove that the link between the overall change in prices and the change in labour force is a deterministic one: inflation is one-to-one function of labour force. Therefore, the Banque de France and other economic/financial authorities are able to control inflation by various tools beyond simple changes in the velocity of circulation of money, as implies the version of monetary policy developed by the ECB (2004). Secondly, the reaction of inflation to the changes in workforce is delayed by five years, and thus, one can obtain an accurate estimate of inflation at a five year horizon. Moreover, various long term projections of labour force, which are based on population projections and estimates of participation in labour force, may provide inflation forecasts at extremely long horizons of 30 to 50 years. Overall, this mid-term (1 to 5 years) and long-term (say, 40 years) forecasts can be considered as "inflation expectations" in the New Keynesian Phillips Curve (NKPC) models. The family of Phillips curves and the concept of inflation as a function of various parameters of economic activity are discussed in Section 2.

The assessment of success in inflation forecasting was dramatically reshaped by Atkeson and Ohanian (2001). They demonstrated that, at a four quarter horizon, a univariate autoregressive model predicts inflation in the USA better than the most elaborated models based on economic and financial variables. This finding became a benchmark for inflation forecasts and just a few models are able to slightly outperform the naïve (no-change) forecast during short periods of time, as demonstrated by Stock and Watson (2006, 2007, 2008). Therefore, a direct comparison of various forecasting models is not necessary any more. One has to estimate the improvement on the naive model. As a measure of the predictive power the root-mean squared forecast error (RMSFE) is often used.

There are two general econometric approaches to inflation forecasting: univariate and multivariate. The univariate models describe inflation as an independent stochastic process, which may include a varying number of autoregressive terms. For aggregated inflation measures, Stock and Watson (2007) introduced a new benchmark univariate model. They estimated various versions of the unobserved components stochastic volatility (UC-SV) model and showed that it is able to outperform the AR(1) model at a few quarter horizon, but not always. As an alternative to the prediction of such aggregated inflation measures as the GDP deflator or consumer price index (CPI), Espasa *et al*. (2002) and Marcellino *et al*. (2003) proposed the aggregation of forecasts of disaggregate inflation components in order to improve the accuracy of short-term forecasts. Hubrich (2005) demonstrated that the models with disaggregate price components do not provide improvement on a one year horizon for the euro zone. Hendry and Hubrich (2010) reported marginal improvements in the accuracy of inflation forecasts and reduction in the estimation uncertainty when a large number of selected disaggregate variables were used in the model for the aggregate inflation measures.

For France, Hall and Jääskelä (2009) investigated statistical properties of the rate of headline CPI inflation from 1977 to 2008 and compared several forecasting models. This study was conducted within a broader approach comparing inflation in developed countries with explicit inflation target (Australia, Canada, New Zealand, Sweden, and the United Kingdom) and non-targeting countries (Austria, France, Germany, Japan, and the United States). To model the evolution of inflation they used the original and modified UC-SV models and found just a slight improvement in the forecasting accuracy compared to the naïve model, which suggests no change over time.

Following the pioneering work of Stock and Watson (1999, 2002), who were the first to apply Principal Component Analysis to large sets of predictors, Bruneau *et al*. (2007) compiled a set of more than 200 macroeconomic variables for the period between 1988 and 2003 and tested their predictive power against a simple autoregressive model. Their modelling and forecasting was carried out within the Phillips curve framework. For the headline inflation, the authors found only slight (~10%) and temporary improvements in the predictive power at a one-year horizon. Celerier (2009) revisited the previous forecasting model used by the Banque de France (Jondeau *et al*., 1999) and presented a new version combining the Phillips curve and the mark-up model for prices. For some exogenous predictors, this new model outperformed by twenty to thirty per cent a standard AR(4), a VAR, and a non-constrained model at time horizons from one month to 6 quarters.

Overall, the literature on inflation forecast for France does not present any significant improvement on the naïve prediction at a one year horizon. In our original paper (Kitov, 2007) we showed the possibility to predict inflation at a two to five year horizon for the period between 1970 and 2004. The accuracy of prediction was improved by a factor of two and more relative to the "no change" version. Since this accuracy depends only on the precision of labour force estimates it can be easily and significantly improved. To a certain extent, labour force projections play the role of inflation anchor in monetary policy since they define "inflation expectations".

The remainder of this paper consists of four Sections and Conclusion. Section 2 briefly describes the Phillips curve framework and introduces a set of linear and lagged relationships between inflation, unemployment, and the change in labour force. It is also describes the Boundary Element Method used to estimate coefficients in these relationships. In Section 3, we discuss various properties of the involved time series, including the descriptive statistics, stationarity, and structural breaks. Section 4 presents revised inflation and unemployment models for France, reports on quantitative/statistical results for two individual and one generalized link between labour force, inflation, and unemployment, and estimates forecasting errors at various horizons. In Section 5, we discuss general consequences of monetary policy in France considering the accuracy and reliability of the estimated relationships.

## 2. The Fisher/Phillips curve framework

Irving Fisher (1926) introduced price inflation as driving the rate of unemployment. He modelled monthly data between 1915 and 1925 using inflation lags up to five months. The inflation and unemployment time series were short and contained higher measurement errors to produce robust statistical estimates of coefficients and lags in the relevant causal relationship. Kitov (2006) estimated a Fisher-style relationship for the USA using observations between 1965 and 2004 and found that the change in unemployment lags behind the change in inflation by 10 quarters. The 40-year period provides good resolution and high statistical reliability of both regression coefficients and the lag. This relationship was successfully tested for the Granger causality (Granger and Newbold, 1974) and cointegration with the Johansen (1988) test. The two-and-a-half year lag implies the only direction of causality. But other countries may demonstrate different lags and order (Kitov and Kitov, 2010).

Phillips (1958) interpreted the link between (wage) inflation and unemployment in the UK in the opposite direction. The original Phillips curve implied a causal and nonlinear link between the rate of change of the nominal wage rate and the contemporary rate of unemployment. He suggested that wages are driven by the change in unemployment rate. The assumption of a causal link worked well for some periods in the UK. When applied to inflation and unemployment measurements in the USA, the PC successfully explained the 1950s. Then, the PC became an indispensable part of macroeconomics which has been extensively used by central banks ever since. The success of the PC did not last long, however, and new data measured in the late 1960s and early 1970 challenged the original version. When modelling inflation and unemployment in Austria, we follow up the original assumption of a causal link between inflation and unemployment to construct an empirical Fisher/Phillips-style curve.

The period of fast inflation growth in the late 1960s and 1970s brought significant changes to the original PC concept. The mainstream theory had to include autoregressive properties of inflation and unemployment in order to explain the observations. For the sake of quantitative precision, the rate of unemployment was replaced by different parameters of economic and financial activity. All in all, the underlying assumption of a causal link between inflation and unemployment was abandoned and replaced by the hypothesis of "rational expectations" (Lucas, 1972, 1973), and later by the concept of "inflation expectations" (Galí and Gertler, 1999). The former approach includes a varying number of past inflation values (autoregressive terms). It was designed to explain inflation persistency during the high-inflation period started in the early 1970s and ended in the mid 1980s.

The concept of inflation expectations surfaced in the late 1990s in order to explain the Great Moderation (Clarida *et al.*, 2000; Cecchetti *et al.*, 2004; Bernanke, 2004) as controlled by monetary and fiscal authorities (Sims, 2007, 2008). The term "New Keynesian Phillips Curve" was introduced in order to bridge this new approach to the original Keynesian framework (Gordon, 2009). The number of defining parameters has dramatically increased in the NKPC (a few autoregressive terms with varying coefficients) relative to the parsimonious Phillips curve. However, both approaches have not been successful in quantitative explanation and prediction of inflation and/or unemployment (*e.g.*, Rudd and Whelan, 2005ab).

Stock and Watson (1999) were outspoken on data and tested a large number of Phillips-curve-based models for predictive power using various parameters of activity (individually and in aggregated form) instead of and together with unemployment. This purely econometric approach did not include extended economic speculations and was aimed at finding technically appropriate predictors. The principal component analysis (Stock and Watson, 2002) was a natural extension to the multi-predictor models and practically ignored any theoretical background. Under the principal component approach, the driving forces of inflation are essentially hidden.

The original Phillips curve for the UK and the Fisher curve, which could be named as an "anti-Phillips curve", both provide solid evidences for the existence of a causal link between inflation and unemployment. The conflict between the directions of causation can be resolved when both variables are driven by a third force with different lags. Depending on which lag is larger inflation may lag behind or lead unemployment. Co-movement is just a degenerate case.

The framework of our study is similar to that introduced and then developed by Stock and Watson (2006, 2007, 2008) for many predictors. They assessed the performance of inflation forecasting in various specifications of the Phillips curve. Their study was forced by the superior forecasting result of a univariate model (naïve prediction) demonstrated by Atkeson and Ohanian (2001). Stock and Watson convincingly demonstrated that neither before the 2007 crisis (2007) nor after the crisis (2010) can the Phillips curve specifications provide long term improvement on the naïve prediction at a one-year horizon.

Following Fisher and Phillips, we do not include autoregressive components in the Phillips curve and estimate two different specifications for inflation:

$$\pi(t) = \alpha + \beta u(t-t_p) + \varepsilon(t) \tag{1}$$

$$\pi(t) = \alpha_1 + \beta_1 l(t-t_1) + \varepsilon_1(t) \tag{2}$$

where $\pi(t)$ is the rate of price inflation at time $t$, $\alpha$ and $\beta$ are empirical coefficients of the Phillips curve with the time lag $t_p$, which can be positive or negative, $u(t)$ is the rate of unemployment, and $\varepsilon(t)$ is the error term, which we minimize by the least squares (OLS) method applied to the integral (cumulative) curves, with the initial and final levels fixed to the observed ones. In (2), $l(t)=dlnLF(t)/dt$ is the rate of change in labour force, $\alpha_1$ and $\beta_1$ are empirical coefficients of the link between inflation and labour force, $t_1$ is the non-negative time lag of inflation, and $\varepsilon_1(t)$ is the model residual.

Then, we represent unemployment as a linear and lagged function of the change rate in labour force:

$$u(t) = \alpha_2 + \beta_2 l(t-t_2) + \varepsilon_2(t) \qquad (3)$$

with the same meaning of the coefficients and the lag as in (2). We finalize the set of causal models with a generalized version:

$$\pi(t) = \alpha_3 + \beta_3 l(t-t_1) + \gamma_3 u(t+t_2-t_1) + \varepsilon_3(t) \qquad (4)$$

Relationships (2) through (4) have been re-estimated with the data for the past eight years and the Boundary Element Method (BEM) instead of standard regression.

The BEM converts ordinary (also partial) differential equations, *e.g.* relationships (2) through (4), into a set of integral equations. The solution of the integral equations for the period between $t_0$ and $t_{01}$ is an exact solution of the original differential equations. For relationship (2):

$$\int_{t_0}^{t_{01}} d[lnP(t)] = \int_{\tau_0}^{\tau_{01}} (\beta_1 d[lnLF(\tau)] + \int_{\tau_0}^{\tau_{01}} \alpha_1 d\tau + \int_{\tau_0}^{\tau_{01}} \varepsilon_1(\tau)d\tau \qquad (5)$$

where $\pi(t)$ is the rate of change in the price level, $P(t)$, $\tau=t-t_1$, $\int_{\tau_0}^{\tau_{01}} \varepsilon_1(\tau)d\tau = 0$. The solution of integral equation (5) is as follows:

$$ln[P(t_{01})/P(t_0)] = \beta_1 ln[LF(\tau_{01})/LF(\tau_0)] + \alpha_1(t_{01}-t_0) + C \qquad (6)$$

where $C$ is the free term ($C=0$), which has to be determined together with coefficients $\alpha_1$ and $\beta_1$ from the boundary conditions: $P(t_0)=P_0$, $P(t_{01})=P_1$, $LF(\tau_0)=LF_0$, and $LF(\tau_{01})=LF_1$. For 1-D problems, we have fixed values as boundary conditions instead of boundary integrals. The number of boundary conditions in (6) is complete for calculation (or quantitative estimation, if there is no analytic solution) of all involved coefficients. Without loss of generality, one can always set $P_0=1.0$ as a boundary condition. The estimated coefficients $\alpha_1^*$ and $\beta_1^*$ entirely define the particular solution of (6):

$$ln[P(t_{01})] = \beta_1^* ln[LF(\tau_0)/LF(\tau_{01})] + \alpha_1^* (\tau_{01}-\tau_0) \qquad (7)$$

at $t_{01}$, as well as over the entire time interval between $t_0$ and $t_{01}$. It is presumed that $LF(t)$ is a discrete function known from measurements.

The estimation of all involved coefficients gives numerical solutions of 2-D and 3-D problems by the BEM in scientific and engineering applications. In this study, the least-squares method is used to estimate the best fit coefficients. Therefore, the residual between observed and predicted curves is minimized in the L2 metrics. For solving problem (7) with an increasing accuracy, one can run over a series of boundary conditions for subsequent years.

From (7), inflation can be exactly predicted at a time horizon $t_1$ and foreseen at longer horizons with various projections of labour force. A linear combination of $ln[LF(t)/LF(t_0)]$ and $(t-t_0)$ defines any particular solution of (2). The rate of price inflation may change only due to the change in labour force. However, the overall price level may grow even when workforce is constant because of $\alpha_1(\tau_{01}-\tau_0)$ term, for $\alpha_1 \neq 0$.

In terms of the boundary elements method, the right hand side of (7) is the particular solution of the (ordinary) differential equation (2). Since $t_1 \geq 0$, the causality principle holds, and the independent function is known before the dependent one. The only principal difference with the standard BEM used in scientific applications is that the solution (7) is not a closed-form or an analytic solution. The solution is the change in labour force in a given country, which may follow a quite exotic and even stochastic trajectory as related to demographic, social, economic, cultural, climatic, and many other conditions.

### 3. Data

The Organization for Economic Cooperation and Development (2013) provides longer time series for the macroeconomic variables used in our modelling: the GDP deflator (from 1971 to 2012), the CPI (1960 to 2012), the level of total and civilian labour force (1956 to 2012), the rate of unemployment (1968 to 2012). The estimates of consumer price inflation are also available from the U.S. Bureau of Labor Statistics (BLS) (1950 to 2012), which also provides the rate of unemployment (1970 to 2012) and the level of labour force (1970 to 2012). The series for CPI and GDP deflator published by the National Institute of Statistics and Economic Studies (INSEE) and Eurostat almost coincide with those provided by the OECD, but start after 1975. We do not use them in this study.

Three different measures of inflation in France are shown in Figure 1: the OECD CPI, the CPI reported by the BLS, and the OECD GDP deflator (DGDP). Two CPI series practically coincide since 1960 and thus should not be modelled separately. The OECD GDP deflator and CPI inflation are also similar but have relatively large discrepancies (a few per cent per year) during several short intervals. The most important difference is observed in the years of peak inflation between 1974 and 1982 – the CPI peaks are observed one to two years before the corresponding peaks in the DGDP. The shape of inflation curves shows significant change over time 0 from the level of 4% per year in the 1960s, through the peak of 14% per year back to 2% to 3% per year. Econometrically, this observation suggests that the inflation time series are not stationary or include structural break as related to artificial and real reasons.

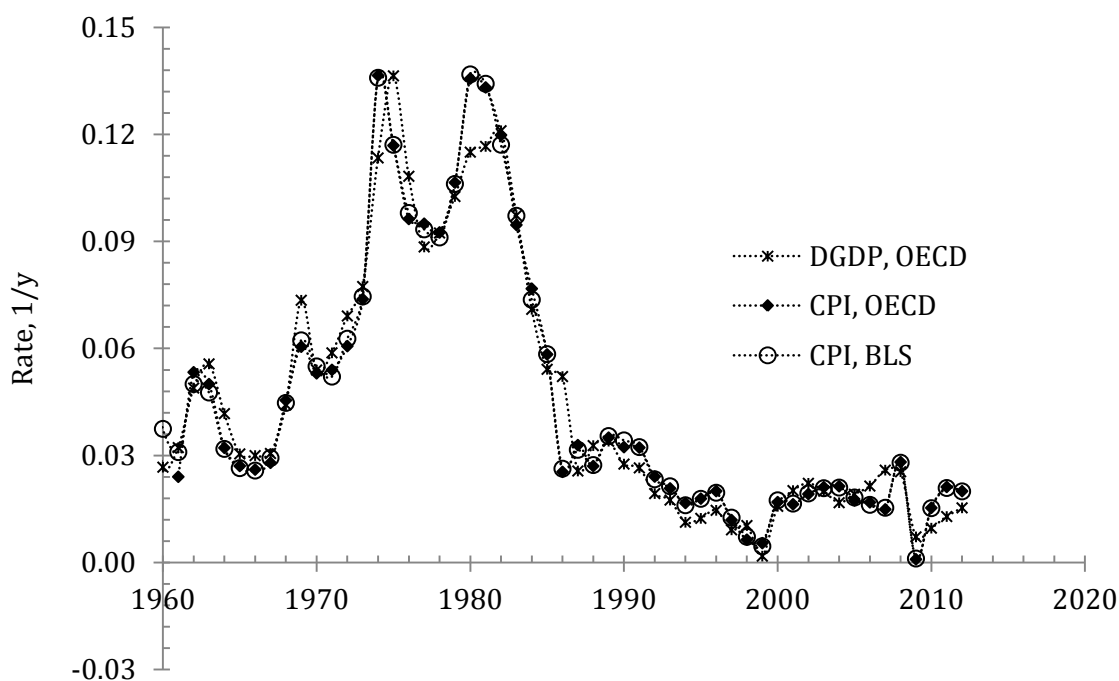

Figure 1. Comparison of three measures of inflation in France – two reported by the OECD and one from the BLS.

We have carried our various unit root tests of the CPI and DGDP time series and their first differences. Table 1 lists select results of the Augmented Dickey-Fuller (ADF) and Phillips-Perron (PP) tests for the period after 1960. Table 1 also presents similar tests for the labour force and unemployment series. The latter is shorter than the other two variables and provides less reliable statistical inferences. Formally, the CPI and DGDP series both contain unit roots and their first differences are I(0) processes. Therefore, inflation process is likely a non-stationary one. However, there are different reasons to consider the results of these unit root tests as biased. Bilke (2005b) tested for stationarity various aggregate and disaggregate inflation time series from 1972 to 2004 and found that the null hypothesis of a unit root can be decisively rejected once one accounts for the structural break in mean. In Figure 1, two distinct periods of elevated (1972 to 1985) and low (1986 to 2012) inflation are clear. Celerier (2009) tested stationarity of several quarterly seasonally adjusted time series (indices) between 1984 and 2008 using the Augmented Dickey-Fuller test and the Ng-Perron test. It was found that most of the tested series are integrated of order one, I(1), processes with a few exclusions of I(2) processes likely associated with the changes in the INSEE statistical methodology or breaks induced by the integration process into the Eurosystem (*e.g.*, new monetary policy). Therefore, the rate of price inflation is considered as a stationary process in our study for the whole period from 1970 to 2012.

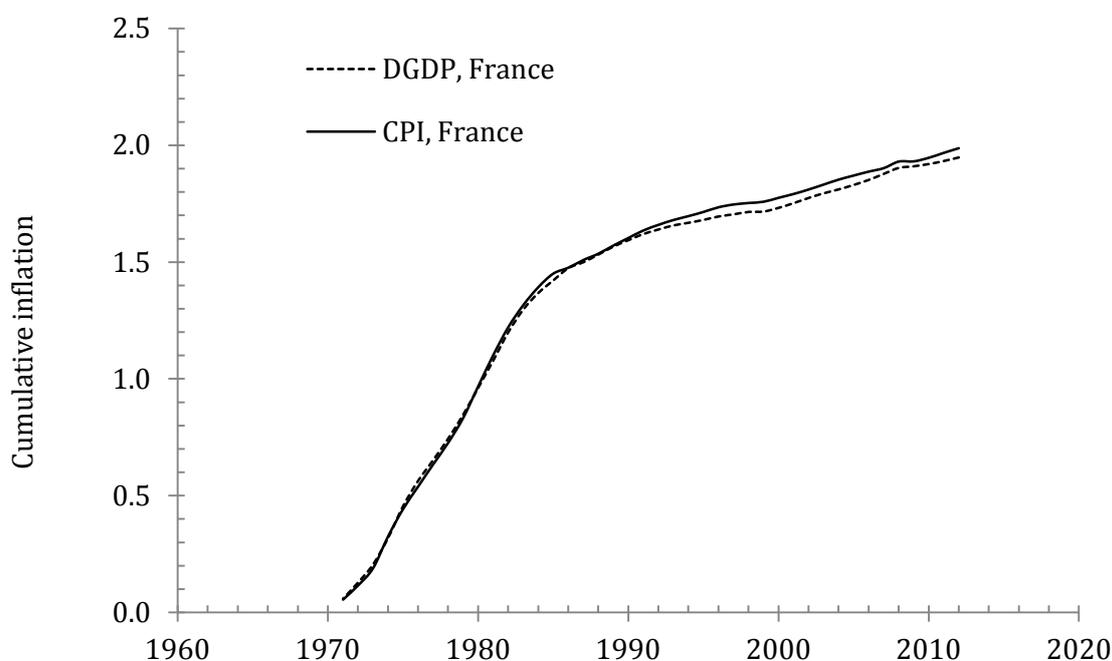

Figure 2. The evolution of the cumulative rate of inflation (the sum of the annual rates of inflation) since 1970. Two measures of inflation rate are compared: the GDP deflator (DGDP) and consumer price index (CPI). The curves start to deviate in 1993.

The structural break in the inflation time series, whether it is artificial or related to monetary policy, deserves a special investigation because of its importance for relationship (2). For example, Figure 2 demonstrates that there was a change in measurement units for the DGDP and CPI – the cumulative inflation curves started to diverge. This change might introduce a break in (2) because the labour force time series is a stationary one (see table 1) and does not reveal any break in the late 1980 and early 1990s. Levin and Piger (2004) found a structural break in the GDP deflator time series in 1993 with the twenty year measurement period started in 1984. This result was partially confirmed by Gadzinski and Orlandi (2004), who allocated breaks in various inflation series in 1992 and 1993. An early break in 1973 was estimated by Corvoisier and Mojon (2004) and Benati (2003). Bilke (2005ab) found one structural break occurred in the mid-eighties using highly disaggregated

CPI representation and interpreted as linked to a major monetary policy change, among several policy related shocks. Overall, Bilke (2005b) distinguishes three possible periods of structural change in French inflation emerge: the early seventies, the mid-eighties, and the early nineties. In this study, we allow one structural break between 1986 and 2003, which is estimated within the BEM with the OLS fit. The estimated structural break is worth to be interpreted in terms of changes in measurements, monetary policy, etc.

Table 1. Unit root tests for the rate of change in labour force and for two measures of inflation and their first differences.

| Test | CPI, OECD[1] | CPI, BLS | DGDP, OECD | u, OECD | u, BLS | l, OECD |
|---|---|---|---|---|---|---|
| **ADF** | -1.45/-6.40* | -1.37/-6.40* | -1.21/-6.26* | -1.35/-4.75* | -1.87/-4.4* | -5.30* |
| **PP z(ρ)** | -5.06/-43.79* | -4.80/-43.03* | -4.09/-41.25* | -3.08/-31.41* | -3.26/-26.71* | -35.53* |
| **PP z(t)** | -1.58/-6.40* | -1.51/-6.30* | -1.39/-6.26* | -1.78/-4.75* | -1.86/-4.36* | -5.28* |

[1] Tests for the original time series / the first difference; * The null of a unit root is rejected for the 1% critical value

The rate of unemployment in France is represented by two time series depicted in Figure 3. Table 1 suggests that both time series are I(1) processes, but the discussion of the inflation measures is likely applicable to $u(t)$. Also notice that the rate of unemployment has been measured since 1968 (OECD) and definitely suffered revision : the unemployment series from 1975 follow the definitions recommended by the International Labour Organization; prior to 1975, the definition of unemployment referred to the number of persons available for work and seeking work (OECD, 2013). Therefore, we model the rate of unemployment as a stationary process since 1970 and include one structural break.

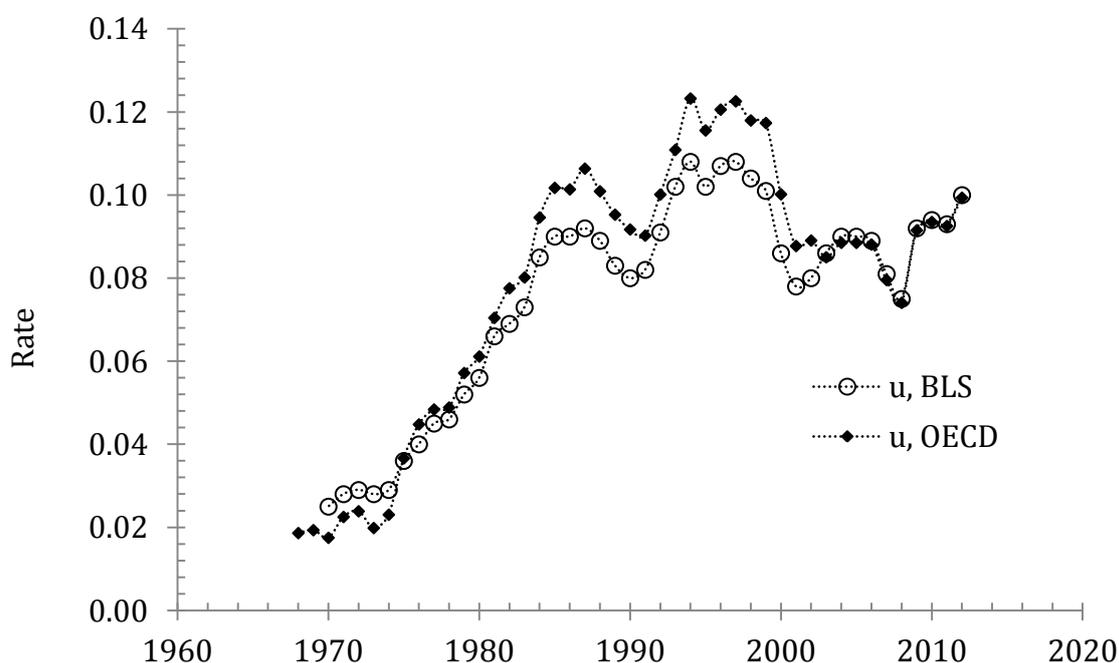

Figure 3. The rate of unemployment in France as reported by the OECD and BLS.

Figure 4 displays the defining variable – the rate of change in labour force, $l=dlnLF/dt$, between 1960 and 2012. There are two series as reported by the OECD and BLS. The difference between two series demonstrates the level of uncertainty in various definitions of labour force. Both series are similar between 1971 and 1984 and then diverge. There are two spikes induced by step revisions to the labour force level, which are likely associated with the corrections to population after decennial censuses.

After 2000, the OECD and BLS estimates of labour force growth are close and almost constant with the mean rate of 0.006 y$^{-1}$ and 0.007 y$^{-1}$, respectively. This period coincides with the Great Moderation when inflation changed in a narrow band between 3% and 1 % per year. Quantitatively, the processes with low signal-to-noise ratio and no-change processes do not provide appropriate dynamic range to estimate a reliable model, and thus, the Great Moderation is a hard period for inflation models (Stock and Watson, 2007). Table 1 evidences that $l(t)$ is a stationary process and can be used for modelling as they are. Two spikes were ironed out - replaced by the average of two neighbouring values.

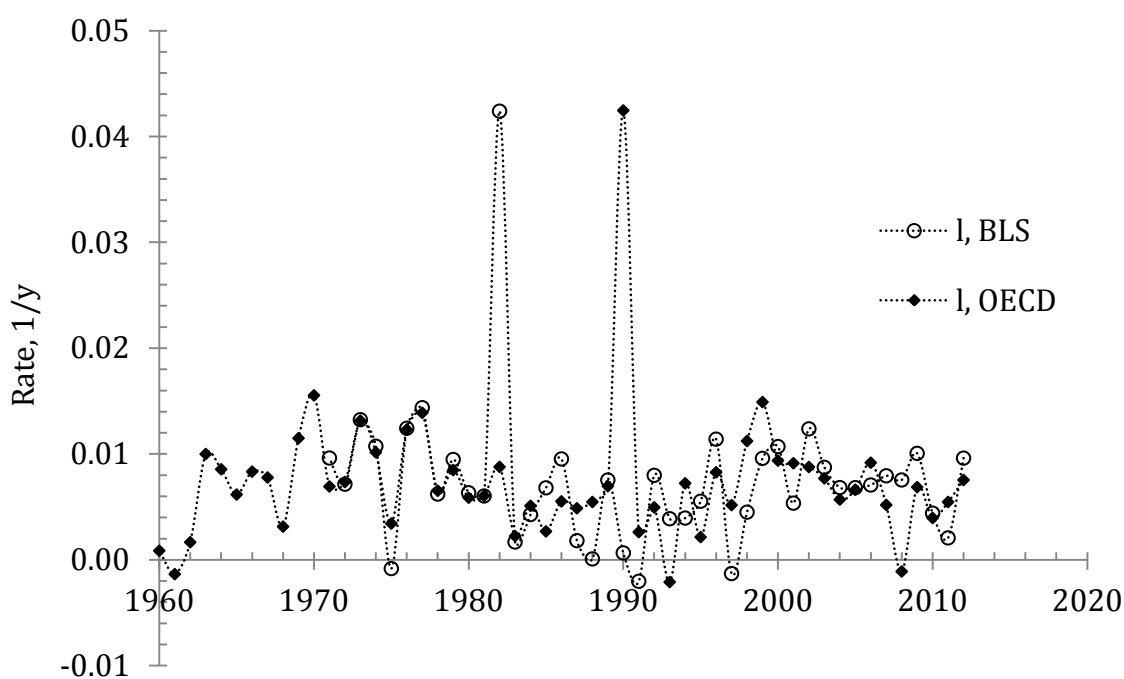

Figure 4. The change rate of labour force reported by the OECD and BLS. Two spikes express corrections in the level of labour force, which were made at different times by the OECD and BLS. For modelling purposes, these spikes were replaced by the average of two neighbouring values.

We are going to model inflation and unemployment in France using the change labour force. The modelled period starts from 1970, i.e. each dependent time series contains 43 readings. It is natural to compare the modelling accuracy with the inherent statistical characteristics of the original time series. Table 2 lists the mean value for each variable together with its standard deviation ($\sigma$). The OECD CPI is characterized by mean of 4.74% per year and $\sigma$=4.07% per year. The DGDP has a lower mean of 3.65% per year and $\sigma$=3.96% per year. These time series are very close in sense of descriptive statistics. However, the first difference of the DGDP has a lower standard deviation (RMSFE1) of 1.29% per year than that of the OECD CPI: RMSFE1=1.56% per year. For the first differences, we use term RMSFE1 instead of standard deviation because it is used as a measure of forecast accuracy for the naive prediction at a one year horizon. The last digit in the RMSFE1 expresses the horizon in years. The lower RMSFE1 for the DGDP suggests that it is smoother than the CPI time series. The estimates of RMSFE1 are the target of forecast precision for inflation models. Table 2 also lists RMSFEs at time horizons from 2 to 5 years, which are used as benchmark prediction accuracy for inflation models in Section 4. The RMSFE doubles with the horizon increasing from 1 to 3 years and rises further for the 5 year horizon.

The growing uncertainty in prediction is one of the reasons for a sound monetary policy to rely on explicit or implicit inflation targeting. Central banks have to aim at the reduction in this uncertainty. The Phillips curve with the change in labour force as the parameter of economic activity allows lowering the uncertainty at a four year horizon by a factor of 3 to 4. The change in labour force is the anchor for inflation expectations sought by central banks.

We have already discussed the possibility of a structural break in the inflation series. It is instructive to estimate RMSFE for two intervals separated by the break. Table 2 lists two standard deviations before and after 1994, which demonstrate large changes in inflation volatility. For the DGDP, st.dev1=1.6 % per year while st.dev2=0.66% per year, i.e. by a factor of 3 lower. Similar fall is observed for two CPI series.

Table 2. Descriptive statistics of the modelled time series and their first differences with varying time lag.

|  | CPI, OECD | DGDP | CPI, BLS | u, OECD |
|---|---|---|---|---|
| **Mean** | 0.0474 | 0.0465 | 0.0475 | 0.0814 |
| **St.dev.** | 0.0407 | 0.0396 | 0.0407 | 0.0302 |
| **RMSFE1**[1] | 0.0156 | 0.0129 | 0.0155 | 0.0072 |
| **St. dev1**[2] | 0.0192 | 0.0163 | 0.0198 | 0.0059 |
| **St. dev2**[3] | 0.0092 | 0.0066 | 0.0089 | 0.0077 |
| **RMSFE2** | 0.0233 | 0.0205 | 0.0233 | … |
| **RMSFE3** | 0.0293 | 0.0257 | 0.0295 | … |
| **RMSFE4** | 0.0334 | 0.0295 | 0.0335 | … |
| **RMSFE5** | 0.0362 | 0.0323 | 0.0360 | … |

[1]Root-mean-squareed forecast error for the naïve forecasting at a one year horizon. The last digit defines the forecasting horizon: from one (RMSFE1) to five (RMSFE5) years; [2] Standard deviation in the first difference for the period between 1970 and 1994; [3] Standard deviation in the first difference for the period between 1995 and 2012.

The average rate of unemployment between 1970 and 2012 is 8.14% with standard deviation of 3%. The unemployment series is much smoother than both inflation series with the RMSFE1=0.7%. This low value reveals little year-on-year changes in unemployment. The change in volatility after 1994 is opposite and much smaller than for the inflation series: standard deviation increases from 0.59% to 0.77%.

### 4. Results

We start modelling with the CPI reported by the OECD. Because of the potential structural break after 1990, we estimated coefficients of the linear and lagged relationship between the CPI inflation and the change in labour force for the period between 1970 and 1990:

$$\pi(t) = 16.0 l(t\text{-}5) - 0.050 \qquad (8)$$
$$\phantom{\pi(t) = }(0.3) \phantom{l(t\text{-}5) - }(0.002)$$

where $l(t\text{-}5)$ is the change rate of the labour force five years before, with the lag is estimated by the model; $p$-values for the slope and intercept are $10^{-22}$ and $2 \cdot 10^{-3}$, respectively. (For other models in this paper, the uncertainty of coefficients and $p$-values are similar and thus omitted.) The upper panel of Figure 5 illustrates the fit between the cumulative observed inflation and that predicted by (8). Notice that the labour force time series starts in 1965 and the prediction of $\pi(t)$ is available through 2017. Therefore, the involved data span the period from 1965 to 2017.

The curves in Figure 5 are close between 1970 and 1995. This is the period when the labour force measurements were not disturbed by the change in monetary policy and step adjustments. The cumulative curves slightly diverge between 1996 and 2002, and the divergence becomes fast after 2002. The consequence of the sought structural break is absolutely clear after 2002, but statistically, it can also be dated earlier. It is not excluded that the models before and after the true break give similar predictions around the break year. Visual timing could be biased and strict statistical estimates are preferred.

There is a five-year lag between the labour force change and the reaction of inflation. The slope of 16.0 indicates that CPI inflation was very sensitive to the labour forced change. The intercept of -0.050 implies that the change rate of labour force must be positive in order to avoid price deflation. The threshold for deflation is the rate of labour force change of 0.0031 $y^{-1}$ (=0.050/16). The actual change rate has been higher than this threshold over the studied period.

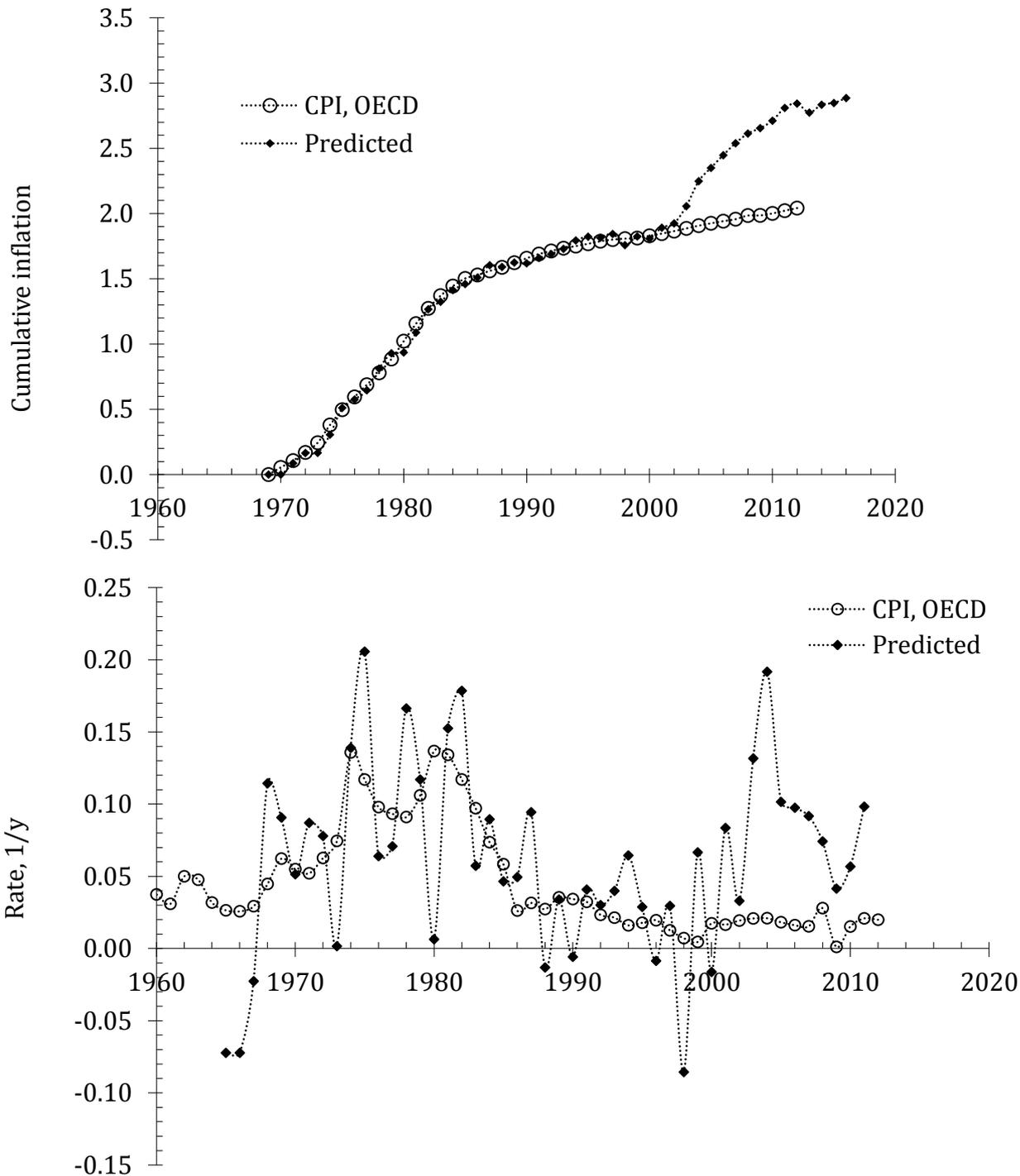

Figure 5. Upper panel: the observed and predicted cumulative inflation, as defined by (8) with the OECD labour force. The DGDP lags by five years behind the labour force change. Notice the discrepancy started in 1999. The rate of predicted inflation oscillates around 10% after 2000. The period after 1999 has to be described by different slope and intercept. Lower panel: the relevant annual curves.

In the lower panel of Figure 5, we depict the annual readings of measured and predicted inflation. Due to the estimated slope of 16.0 the measurement errors in the labour force produce larger deviations between the curves. These residuals are easy to reduce, however by smoothing the predictor. Since the change in labour force leads inflation by 5 years the moving average would reduce this forecasting horizon by a half of averaging period. For a three year moving average of the predictor, $l_3(t-5)$, the forecasting horizon is 4 years.

Now, we extend the modelling period to 2012 and apply the BEM with the OLS fit to the OECD CPI data and the $l_3(t-5)$ series as a predictor. The best fit model is as follows:

$$\pi(t) = 16.108 l_3(t-5) - 0.0500; \quad t \leq 1993$$
$$\pi(t) = 0.952 l_3(t-5) - 0.0093; \quad t \geq 1994 \quad (9)$$

where the time lag is 5 years and the break moves to 1993. The slopes, intercepts, break years, and the coefficients of determination for all models are listed in Table 3. For this particular model, the slop falls from 16 to ~1, and the intercept rises from -0.05 to -0.01. Figure 6 displays the predicted inflation together with the measured one also smoothed with MA(3). The agreement between the curves is good. The model residuals between 1970 and 1990 fall by at least a factor of 2. The labour force time series has relatively large measurement annual errors, but the level is well controlled over longer periods.

The change in break year for the CPI models with various predictors, from $l(t-5)$ to the $l_7(t-5)$ (seven year moving average of $l(t-5)$) is limited to 1993 and 1994 (see Table 3). We conclude that the data provide a reliable estimate of the break. The slope changes from 16.1 to 17.3 for the period between 1970 and 1994. The intercept decreases from -0.5 to -0.6 for two extreme slope estimates. These are rather statistical variations which do not influence the overall fit between the cumulative curves, as reveal the estimates of $R^2$ in Table 3. The slopes and intercepts for the second period are less reliable since the rate of inflation is rather constant since 1995. However, all models with smoothed predictors are characterized by similar slopes (~1.0) and intercepts (~0.01).

Table 3. Coefficients, lags, break years, and $R^2$ for three variables and various predictors

| Variable | Predictor | Slope1 | Intercept1 | Slope2 | Intercept2 | Lag, y | Break | $R^2$ adj., A | $R^2$ adj., C |
|---|---|---|---|---|---|---|---|---|---|
| CPI | $l(t-5)$ | 16.304 | -0.0513 | 2.046 | -0.0001 | 5 | 1994 | 0.5251 | 0.9978 |
| CPI | $l_3(t-5)$ | 16.108 | -0.0500 | 0.952 | 0.0093 | 4 | 1993 | 0.8626 | 0.9993 |
| CPI | $l_5(t-5)$ | 16.363 | -0.0525 | 0.879 | 0.0100 | 3 | 1994 | 0.8550 | 0.9992 |
| CPI | $l_7(t-5)$ | 17.324 | -0.0595 | 1.039 | 0.0089 | 2 | 1994 | 0.8659 | 0.9991 |
| DGDP | $l(t-5)$ | 16.543 | -0.0539 | 2.920 | -0.0065 | 5 | 1994 | 0.5532 | 0.9980 |
| DGDP | $l_3(t-5)$ | 16.348 | -0.0526 | 1.941 | 0.0020 | 4 | 1993 | 0.9212 | 0.9996 |
| DGDP | $l_5(t-5)$ | 16.031 | -0.0501 | 1.574 | 0.0054 | 3 | 1992 | 0.8870 | 0.9996 |
| DGDP | $l_7(t-5)$ | 17.344 | -0.0602 | 1.945 | 0.0027 | 2 | 1993 | 0.9031 | 0.9995 |
| $u$ | $l_3(t-0)$ | -13.684 | 0.1661 | 3.578 | 0.0659 | 0 | 1995 | 0.7817 | 0.9996 |

Figure 7 displays two cumulative curves: for the measured inflation and that predicted by (9). The difference between these cumulative curves was minimized in the OLS sense in order to estimate all parameters in (9). Visually, they practically coincide and the latter actually represents the change in labour force. The adjusted coefficient of determination $R^2$=0.9993 (see Table 3). This estimate of $R^2$ is not biased only when the cumulative curves are cointegrated. We prove that these cumulative curves are cointegrated and provide select results of econometric tests later on. Testing for cointegration extends the causal link between inflation and labour force to the level of vector autoregressive (VAR) representation, which needs additional consideration. Here, we just stress that this coefficient of determination is not biased.

In the long run, one can replace the growth in cumulative inflation (this is not the consumer price index) by the change in the labour force with increasing precision. The annual residuals will be retained at the same level, however, due to measurement errors. However, the coefficient of determination for the annual curves is extremely high for a model without autoregressive terms: $R^2=0.855$. This means that eighty six per cent of variability in the rate of CPI inflation is explained by the change in labour force at a four year horizon. For the original predictor, $l(t-5)$, $R^2=0.52$, as one could expect from larger fluctuations in the predicted series.

Within our framework, the residual difference between the observed and predicted readings is related to measurement errors. In France, the level of labour force is measured with an uncertainty, which is not appropriate for modelling of likely more accurately measured inflation. One year measuring baseline is not enough for obtaining reliable estimates of the change rate of labour force. Moving average (or other low-pass filter) takes an advantage of a longer baseline for the calculation of the change rate, and thus, may provide a significant increase in the predictive power. Therefore, a longer time unit will potentially result in a higher accuracy of corresponding measurements and in a better correlation between all modelled variables. Table 3 shows that smoothing with moving average can significantly increase the coefficient of determination with a small decrease in forecasting horizon. Broadening of averaging windows beyond 3 years does not increase the agreement, however, and the optimal forecasting horizon is likely 4 years. One does not need to reduce the horizon in expense of lower resolution.

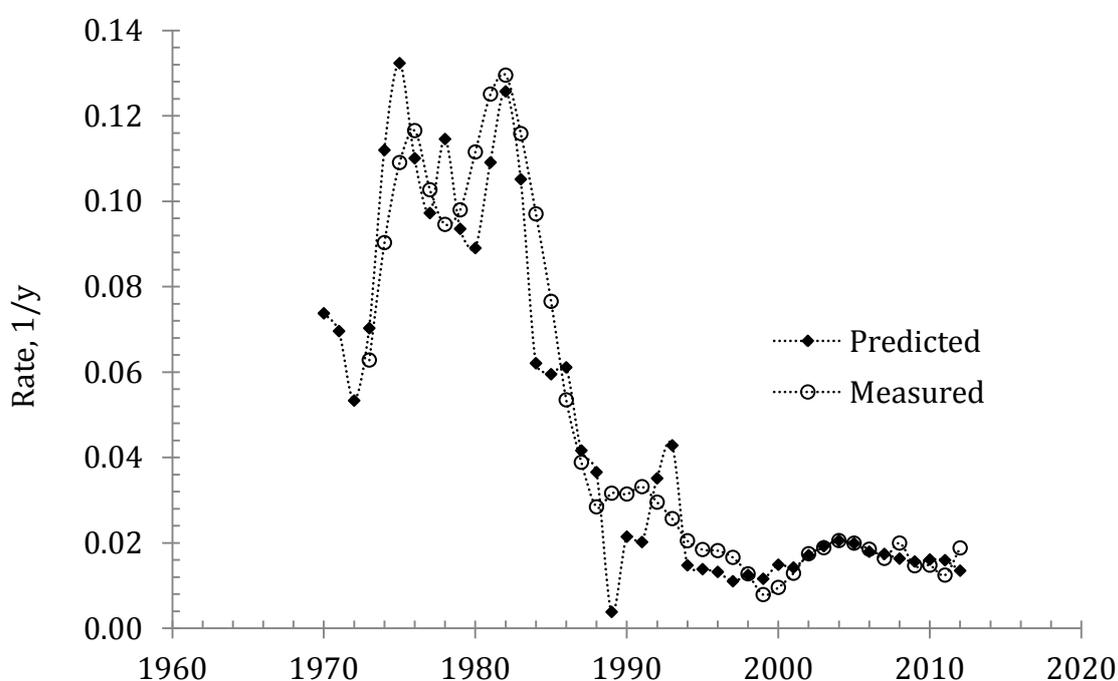

Figure 6. The observed CPI inflation smoothed with MA(3) and that predicted by (9) with $l_3(t-5)$ as a predictor (see Table 3); $R^2=0.91$. For the original CPI series with the same predictor, $R^2=0.86$.

Model (9) and other CPI models in Table 3 are reliable and provide excellent agreement between measured and predicted time series since 1970 for the annual and cumulative versions. Table 4 lists the relevant root-mean square forecast errors. For the basic model with $l(t-5)$ as a predictor, RMSFE=0.035 $y^{-1}$ at a 5 year horizon. This is marginally better than RMSFE5=0.036 $y^{-1}$ estimated by the naïve model at the same horizon. The improvement in accuracy is negligible ~3%. For the smoothed predictors, RMSFE decreases for our model as well as for the naïve prediction with decreasing forecast horizon. The gain in predictive power associated with the smoothed predictor is significant. It is the highest for the $l_3(t-5)$ version – 56% ((0.034-0.015)/0.034). In other words, the RMSFE of our model is by a factor of 2.3 lower than the naïve RMSFE4. For a three year horizon ($l_5(t-5)$) the gain is 48%, and for a two year horizon ($l_7(t-5)$) – only 35%. There is no model in the literature on inflation forecasting in France or any other country which demonstrates such an excellent performance at 2 to 4 year horizons. It is worth noting that the estimates of RMSFE for the cumulative curves are very close to those from the annual series.

Table 4. RMSFEs at various time horizons and time intervals for the OECD CPI and DGDP.

| Predictor | Horizon, y | CPI | sd1/sd2[1] | VECM[2] | DGDP | VECM | sd1/sd2 |
|---|---|---|---|---|---|---|---|
| $l(t-5)$, A[3] | 5 | 0.035 | 0.048/0.008 | ... | 0.035 | ... | 0.045/0.011 |
| $l(t-5)$, C[4] | 5 | 0.028 | 0.036/0.009 | 0.015 | 0.026 | 0.012 | 0.032/0.014 |
| Naive[5] | 5 | 0.036 | 0.047/0.010 | ... | 0.032 | ... | 0.043/0.009 |
| $l_3(t-5)$[6], A | 4 | 0.015 | 0.020/0.008 | ... | 0.010 | ... | 0.013/0.005 |
| $l_3(t-5)$, C | 4 | 0.020 | 0.026/0.006 | 0.012 | 0.012 | 0.008 | 0.014/0.005 |
| Naive | 4 | 0.034 | 0.044/0.009 | ... | 0.029 | ... | 0.039/0.010 |
| $l_5(t-5)$, A | 3 | 0.015 | 0.028/0.005 | ... | 0.013 | ... | 0.018/0.005 |
| $l_5(t-5)$, C | 3 | 0.023 | 0.028/0.005 | 0.010 | 0.015 | 0.007 | 0.018/0.005 |
| Naive | 3 | 0.029 | 0.038/0.010 | … | 0.026 | … | 0.034/0.010 |
| $l_7(t-5)$, A | 2 | 0.015 | 0.019/0.006 | ... | 0.012 | ... | 0.016/0.005 |
| $l_7(t-5)$, C | 2 | 0.023 | 0.028/0.005 | 0.009 | 0.016 | 0.005 | 0.019/0.004 |
| Naive | 2 | 0.023 | 0.030/0.009 | … | 0.021 | … | 0.027/0.009 |

[1] sd1/sd2 are RMSFEs for the periods between 1970 and 1994 and 1995 and 2012; [2] VECM specifications: rank=1, max lag in VAR = 4, trend = none; [3] Annual residuals; [4] Cumulative residuals; [5] Standard deviations for the naïve prediction at a given horizon; [6] $l_3(t-5)$ is the three year moving average (MA(3)) of $l(t-5)$

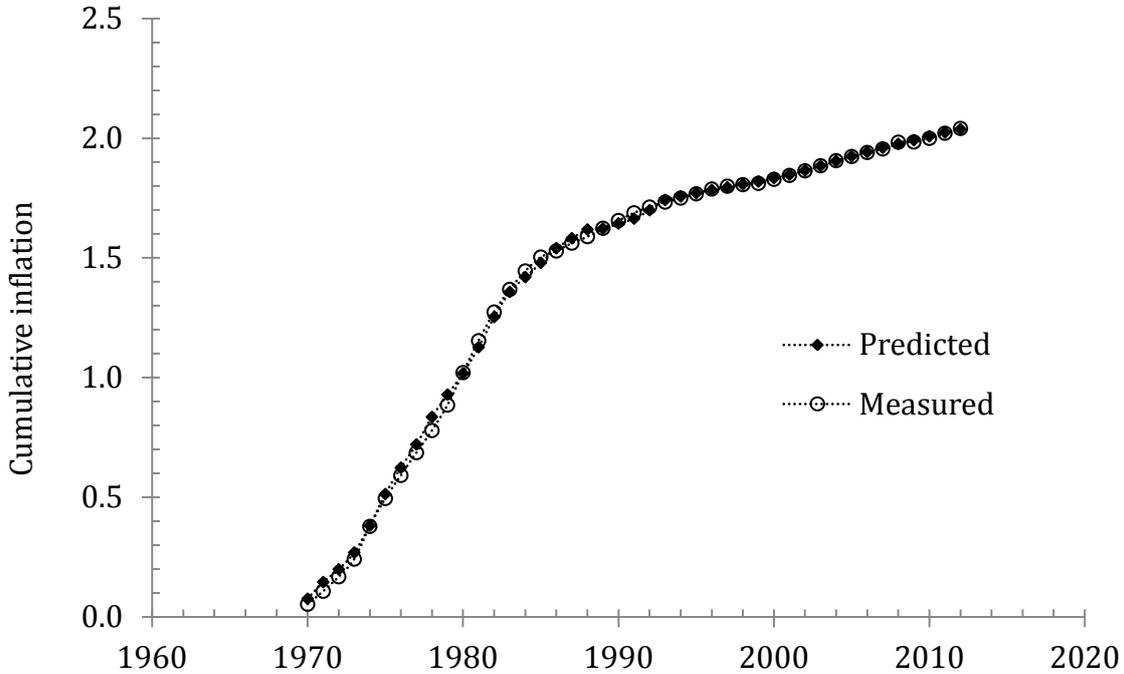

Figure 7. The observed and predicted cumulative CPI inflation for the $l_3(t-5)$ (the centred three-year moving average of $l(t-5)$) as a predictor. RMSFE=0.015 at a four year horizon; $R^2$=0.9993.

There is a structural break in (9), and thus, one has to estimate RMSFE for two intervals separately. Table 4 provides this information. For the $l_3(t-5)$ predictor, the improvement in RMSFE is larger during the initial period. The naïve prediction at a four year horizon since 1994 gives an extremely low RMSFE=0.009 $y^{-1}$, which manifests low-amplitude fluctuations of the CPI inflation around 2% per year with only two excursions in 1999-2000 and 2008-2009 (see Figure 1). The "no change" processes are the hardest to model since they do not provide appropriate dynamic range and higher signal-to-noise ratio. Therefore, the influence of measurement noise increases dramatically and spoils model predictions. Nevertheless, our model provides an improvement in the RMSFE for the second interval by a factor of 2 at a three year horizon with $l_5(t-5)$ as a predictor. The model has a good predictive power since 1995 because the smoothed change in labour force is almost constant during this period.

The DGDP deflator covers the change in all prices in a given economy, and thus, is characterized by lower fluctuations over time than the CPI (see Table 2). As before, we apply the BEM and the OLS to the original time series as well as to the predictor smoothed with moving average. Figure 8 and Tables 3 and 4 present select results for the OECD GDP deflator. The best fit relationships for the change rate in labour force are as follows:

$$\pi(t) = 16.54\, l(t\text{-}5) - 0.0539;\ t<1995$$
$$\pi(t) = 2.92\, l(t\text{-}5) - 0.0065;\ t>1994 \qquad (10)$$

where the breaking year is 1994, and the relevant coefficients and their change are similar to those for the CPI. The coefficient of determination $R^2$=0.55, i.e. just marginally higher than for the CPI. The estimated lag allows forecast at a five year horizon. Figure 8 demonstrates that the rate of inflation in France should be very low if not negative in 2013 and then will return to the level of 1% per year. Inflation if France is well anchored by the low growth rate in labour force and the current version of monetary policy.

The estimated RMSFE=0.035 $y^{-1}$ at a five year horizon is the same as obtained for the CPI. However, the naïve prediction gives a lower RMSFE=0.032 $y^{-1}$ for the DGDP. Our model for the DGDP is poor at a five year horizon. When smoothed with MA(3), the original predictor provides a significant improvement on the "no change" prediction at a four year horizon: RMSFEs are 0.010 $y^{-1}$ and 0.029 $y^{-1}$, respectively. The gain is 66%. For wider averaging windows the gain falls to approximately 50%, i.e. our models halve the naïve RMSFEs. When estimated for two intervals before and after the break in 1994, the RMSFE practically repeat the pattern reported for the CPI. The highest gain is obtained at a four year horizon with $l_3(t\text{-}5)$ as a predictor: the relevant RMSFEs are by a factor of 3 lower for the earlier period and halved for the period since 1995.

Formally, two non-stationary processes, like the predicted and measured cumulative inflation, may produce spurious regression (Granger and Newbold, 1974), which biases the coefficient of determination up. The problem of spurious regression was successfully resolved in econometrics and several statistical tests were developed. At this stage, we are interested to prove that the estimates of $R^2$ for the cumulative curves in Table 3 are not biased. The Johansen tests (see Table 5) result in rank 1 for all versions of the cumulative predicted curves for both CPI and DGDP. Therefore, the measured and predicted curves are cointegrated when the VAR representation is applied.

Table 5. Unit root tests of the annual and cumulative model residuals: Cointegration ADF (CADF) test and Johansen test for cointegration.

| Test | CPI, $l$ | CPI, $l_3(t\text{-}5)$ | DGDP, $l$ | DGDP, $l_3(t\text{-}5)$ |
|---|---|---|---|---|
| | | **CADF** | | |
| **ADF_A** | -7.67* | -5.02* | -8.16* | -5.48* |
| **ADF_CU** | -5.28* | -2.71 | -5.68* | -3.40 |
| **PP_A** | -37.44** | -27.16** | -30.72** | -26.34** |
| **PP_CU** | -30.08** | -13.43 | -32.57** | -15.32 |
| | | **Johansen test** | | |
| **Rank** | 1 | 1 | 1 | 1 |
| **Trace statistics** | 0.11*** | 0.19*** | 0.38*** | 0.097*** |
| **Eigenvalue** | 0.54 | 0.47 | 0.51 | 0.56 |

\* The null of the unit root is rejected for 1% critical value (-4.32) defined by Engle and Granger (1987)
\*\* The null of the unit root is rejected for 1% critical value
\*\*\* Trace statistics for rank 1

The CADF test proposed by Engle and Granger (1987) shows that the difference between the cumulative curves is an I(0) process for $l(t\text{-}5)$ used as a predictor (Table 5). For the $l_3(t\text{-}5)$ as a predictor, the residual of the cumulative curves has a clear periodic component with higher autocorrelation, i.e. this predictor does not produce an I(0) residual time series despite the fact that the standard error is reduced by a factor of 2, as Table 3 shows. Figure 9 depicts the model residuals for the annual and cumulative curves for both predictors. The ADF and PP unit root tests reveal that the null of unit root in the cumulative model residuals for the $l_3(t-5)$ are not rejected. The prediction with $l_3(t-5)$ is much better in terms of measurements, but has poor statistical properties. The Johansen test is more reliable than the CADF since it resolves the problems of autocorrelation and heteroscedasticity in the residual time series. We consider the measured and predicted cumulative curves as cointegrated and thus the estimates of $R^2 \sim 1.0$ in Table 3 are not biased.

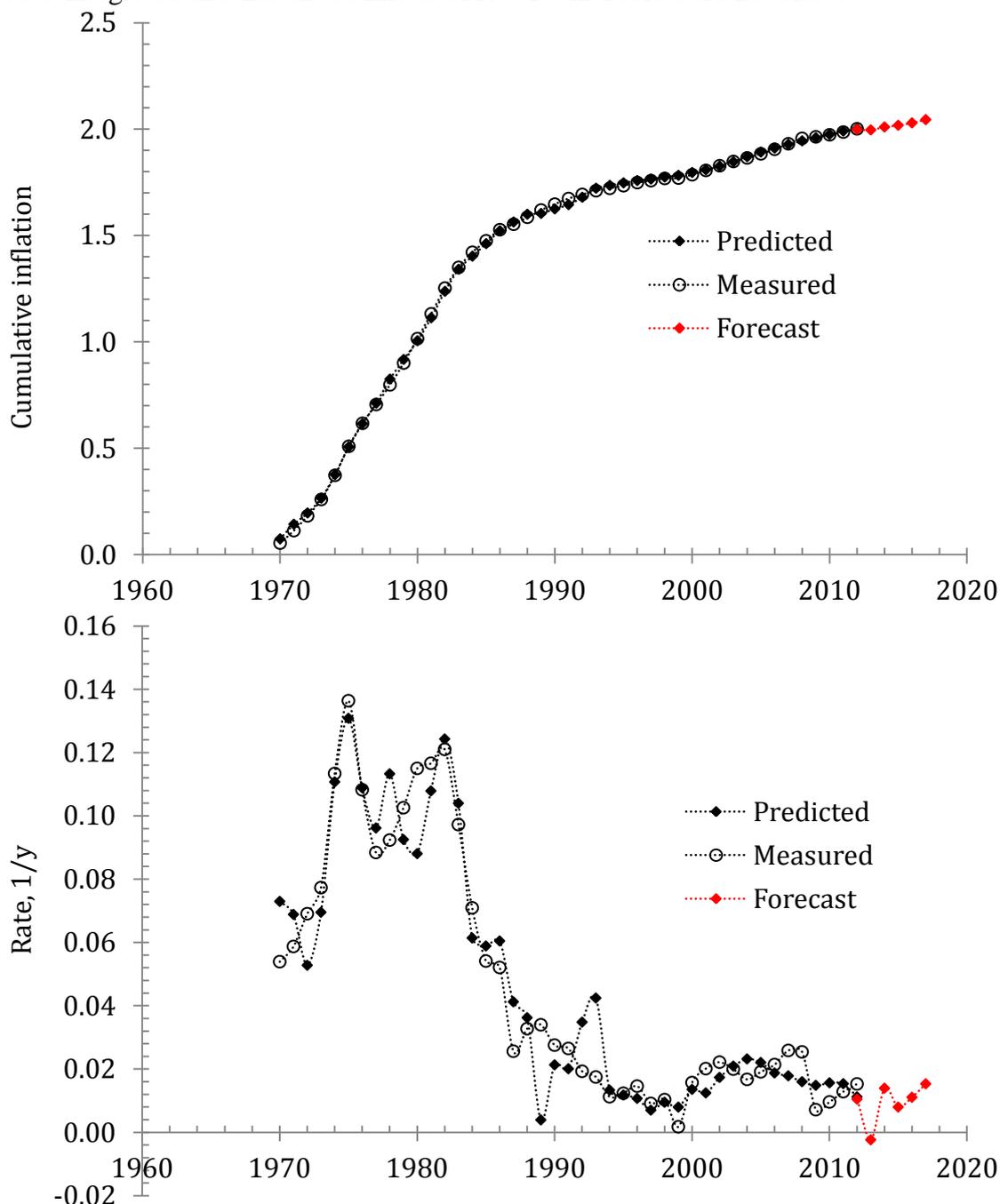

Figure 8. The measured and predicted rate of inflation represented by the GDP deflator. Red line is a five year forecast. Upper panel: the cumulative curves. Lower panel: the annual predicted time series smoothed with MA(3). In 2013, the rate of inflation will be very low or negative, and then will return at the level of ~1.0% per year.

We have been discussing static relationship (8) between the cumulative inflation, *lnP*, and the cumulative rate of change in labour force plus the input of linear time trend. There exists a causal link between these variables, with the change in labour force being a stochastic process. Within the realm of physics, the strategy of making this link more reliable consists in more accurate measurements. For a causal link, the cumulative variables should coincide when they are measured precisely. However, no improvement is possible for the past values. In terms of econometrics, one has an opportunity to improve the model standard error by using statistical properties of the residual time series.

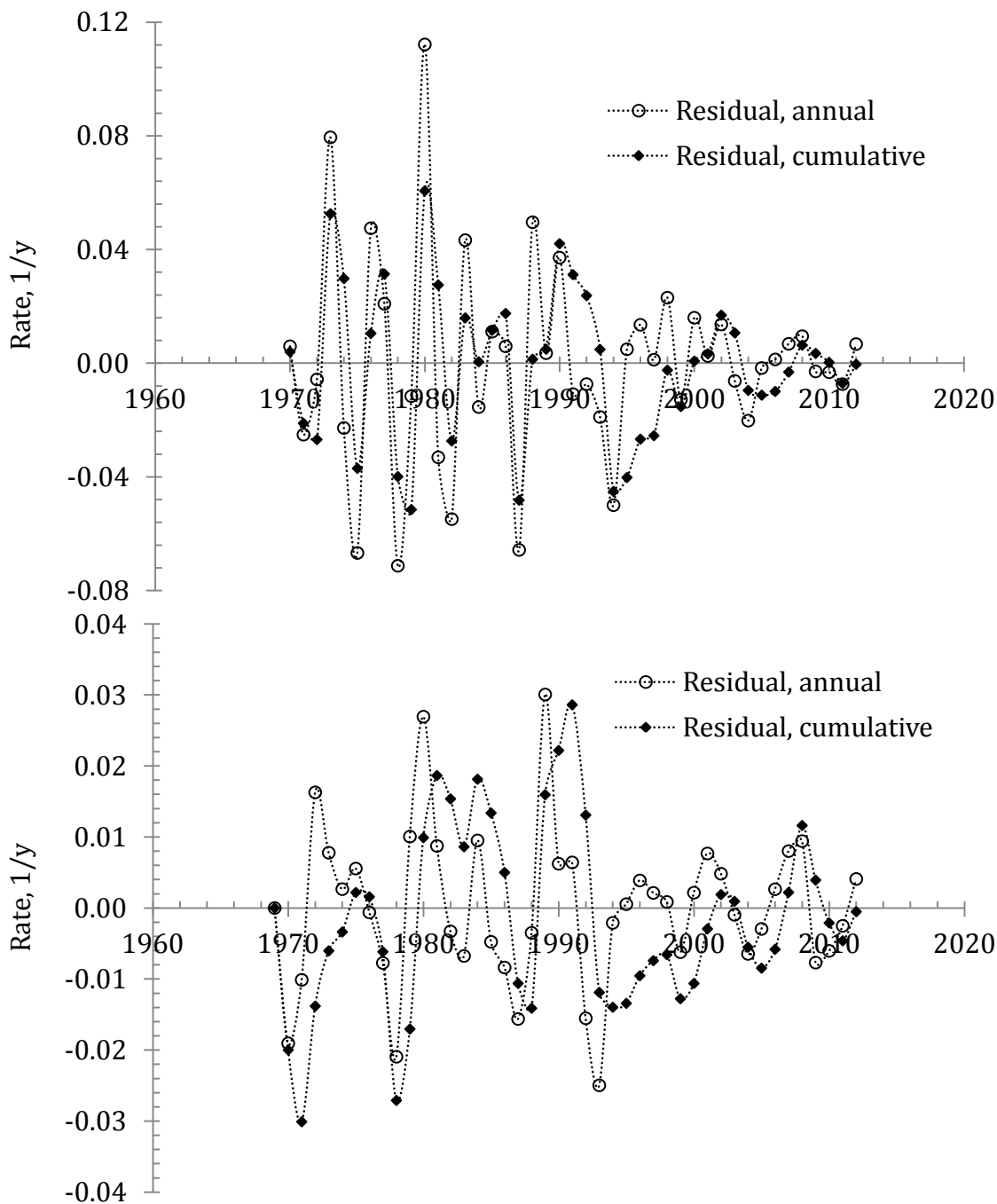

Figure 9. The model residuals for the GDP deflator. Upper panel: predictor – $l(t-5)$. Lower panel: predictor - $l_3(t-5)$.

Testing the measured and predicted cumulative inflation, which are both I(1) processes, for cointegration one can represent the error term using an autoregressive distributed lag model. In the simplest dynamic autoregressive form with one lag term:

$$P(t) = \lambda_1 P(t-1) + \lambda_2 X(t) + \lambda_3 X(t-1) + e(t) \tag{11}$$

where $P(t)$ is the cumulative inflation at time $t$, $X(t)$ is the predicted cumulative value, $\lambda_i$ are empirical coefficients, $e(t)$ is the autoregressive model error. Since we have applied the BEM to predict inflation, $\lambda_2=1$. In this representation, $\lambda_1 P(t-1) + \lambda_3 X(t-1) + e(t) = \varepsilon_1(t)$ and the LHS may have predictable components and thus improve statistical properties of $e(t)$ relative to $\varepsilon_1(t)$. For perfect measurements and the causal relationship $P(t)=X(t)$, $\lambda_1 P(t-1) + \lambda_3 X(t-1) = 0$. Therefore, (11) suggests the presence of measurement errors and/or stochastic trends with error correction (Hendry and Juselius, 2001).

In the error correction modification, (7) is embedded into (11) (Hendry and Juselius, 2000). The simplest notation includes only one lagged value of predictor:

$$P(t) - P(t-1) = \Gamma_1[X(t) - X(t-1)] - \Gamma_2[P(t-1) - X(t-1)] + \upsilon(t) \tag{12}$$

where $\Gamma_1$ is the coefficient of the casual link between $P(t)$ and $X(t)$, which represents the steady-state solution and fixed to 1 in this study, $\Gamma_2$ is the coefficient defining the speed of adjustment towards this steady-state, and $\upsilon(t)$ is the error term in the (vector) error correction model. The importance of term $\Gamma_2 [P(t-1) - X(t-1)]$ is defined by the properties of measurement errors in both variables. For example, the positive correction to the OECD labour force in 1990 (Figure 3) had to compensate the population, and thus, labour force undercounting during the previous period. Such corrections should be evenly distributed over the whole relevant period that had not been done by the OECD. This undercounting induces the discrepancy between $P(t)$ and $X(t)$ during this period and introduces auto-correlated errors in (10). The VECM should suppress the influence of such measurement errors and improve statistical properties of $\upsilon(t)$ relative to $\varepsilon_1(t)$ as well as the influence of other non-stochastic factors.

Model (12) for the link between the cumulative measured and predicted inflation explains the growth rates in the total inflation, $P(t)$, by the sum of the growth in the rate of labour force and linear trend, $X(t)$. In the original VECM representation, the maximum lag can be fixed or estimated from data. We consider a VECM with the following specifications: rank=1 (two variables), maximum lag 4, and "no trend" specification since any possible trend is already compensated in (8).

The VECM does contribute to the improvement in the model residual. Table 4 lists VECM results for the CPI and DGDP time series with the $l(t-5)$ and its 3-, 5-, and 7-year moving averages as predictors. For the DGDP, model (10) outperforms the naïve prediction at a four year horizon by a factor of 2.4 (0.029/0.012). The VECM reduces the standard error from 0.012 $y^{-1}$ to 0.008 $y^{-1}$, i.e. by ~30%. Altogether, the RMSFE is by a factor of 3.6 lower than that of the naive forecast. The same factor is observed for a 3- and 2-year horizon, with the minimum RMSFE of 0.005 $y^{-1}$ for $l_7(t-5)$ as predictor. All in all, the VECM is useful for forecasting purposes but hardly change the reliability of the causal link between inflation and labour force, which is characterized by $R^2=0.999$ and cointegration between measured and predicted series.

Having a few excellent models for inflation we now predict the rate of unemployment with the same tools – the BEM and OLS with a structural break between 1986 and 2003. The predicted and measured, annual and cumulative, curves for the OECD unemployment rate between 1970 and 2012 are presented in Figure 10. To model the period after 1970 is also in line with many other studies devoted to various modifications of the Phillips curves in European countries. The period before 1970 is rarely covered (*e.g.*, Angelini *et al.*, 2001; Canova, 2002, 2007; Espasa *et al.*, 2002; Gali *et al.*, 2001; Marcellino *et al.*, 2003). The best fit model is as follows:

$$u(t) = -13.684 l_3(t-0) + 0.166; \quad 1970 \leq t \leq 1995$$

$$u(t) = 3.578 l_3(t-0) + 0.066; \quad t \geq 1996 \qquad (13)$$

where $l_3(t-0)$ is the three year moving average, the lag $t_2=0$ years, the break is 1995 (likely associated with the introduction of new monetary policy), and the slope changes from -13.7 to +3.6. Both intercepts are positive and imply large unemployment (~17% before 1995 and ~7% after 1995) in the absence of labour force growth, $l(t)=0$. Before 1995, the slope in (13) amplifies the change in labour force, and thus, all measurement errors by a factor of 13.7. This coefficient is also a negative one, i.e. any increase in labour force was converted into a synchronized drop in the rate of unemployment in France. After the new monetary policy was introduced, the growth in labour force pushes unemployment up.

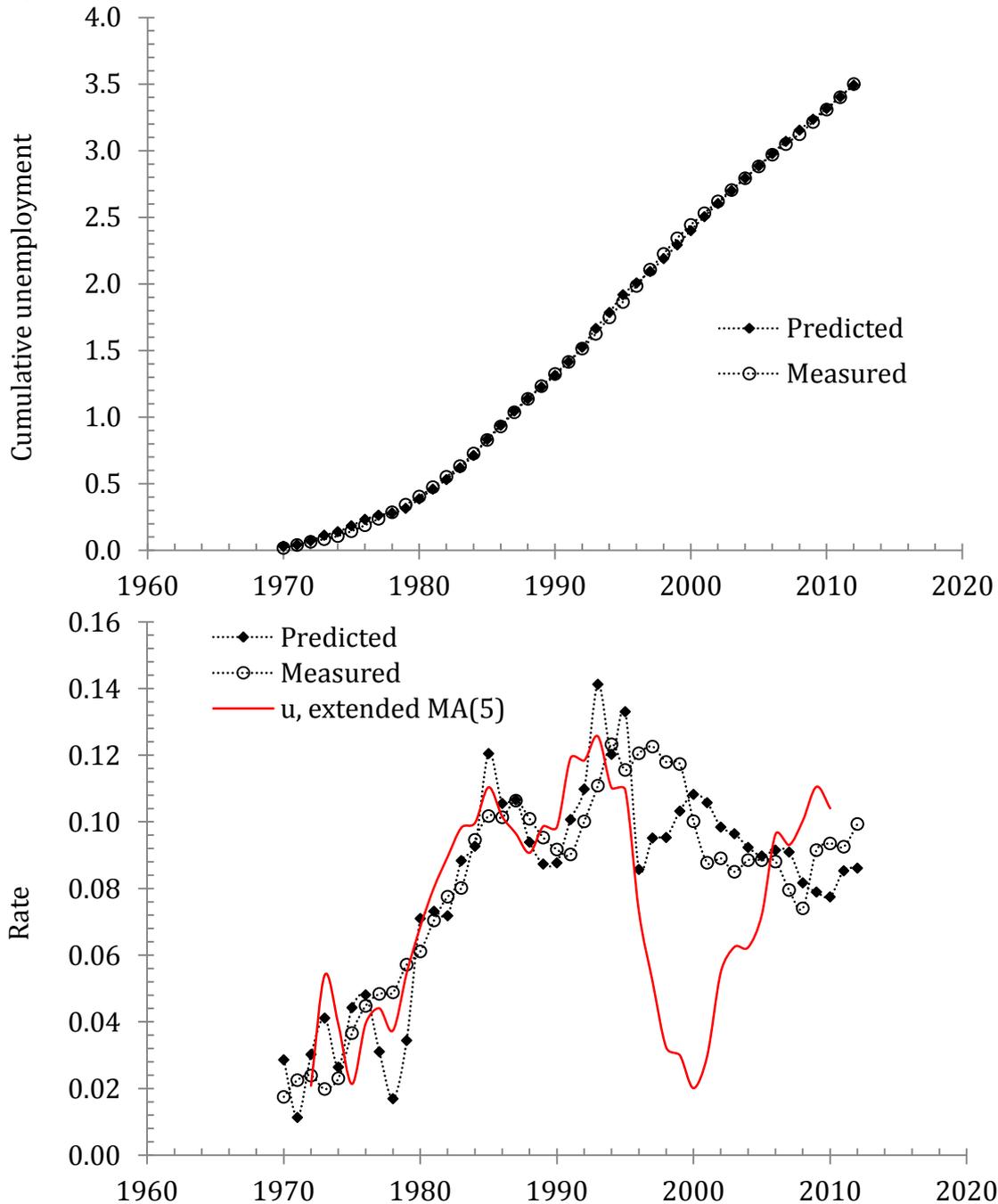

Figure 10. Comparison of observed and predicted unemployment in France. Upper panel: the measured cumulative curve and that predicted by (13). Lower panel: the annual measured and predicted values of unemployment since 1970. Red line represents the earlier relationship smoothed with the MA(5) and extended after 1995.

From 1970 through 2012, there is a good agreement between the observed and predicted curves with $R^2$=0.78 for the annual curves and $R^2$=0.9996 for the cumulative ones. We do not test the cumulative curves for cointegration because the time series are too short and the test results would not be reliable. However, the fit between the annual unemployment curves, which are likely I(0) processes, clearly demonstrates that the level of correlation between the cumulative time series is not biased up. The observed unemployment curve gradually rises from 3% in 1970 to almost 10% in 2004, falls to the level of 7% in 2008, and has been increasing since then. The predicted curve fluctuates around the observed one with an amplitude reaching 0.02. In 1995, a sudden drop in the predicted curve manifests the start of a major deviation from the measured curve. The predicted curve falls from 0.1 in 1994 to 0.02 in 1999.

In Figure 10, we extended the (smoothed with MA(5)) relationship estimated before 1995 into the later period. The predicted curve falls to 0.02 in 1999 unlike the observed one, which hovers around 0.1 from 1985 to 2012. In terms of measuring units or/and economics, this is a structural break, which needs a formal introduction of different relationships before and after 1995. The year of 1995 is definitely a breaking point. There are two potential explanations of the deviation. One is associated with the change in units of measurements, as has been found for Austria (Kitov, 2013). Considering the difference between revisions to the involved data series reported by the OECD and BLS this assumption cannot be ruled out despite the change in the labour force and unemployment definitions is not well documented. Another possibility is that the slope and intercept in (13) were changed in 1995 by some external forces, but the linear link with the labour force was retained. This assumption is appropriate for the rate of inflation suffering a structural break at the same time. The generalized relationship (4) has to take care about such structural breaks in individual components, and thus, to replace (2) and (3). We examine this assumption in detail. Both explanations can be expressed as the change of coefficients in (13). It is important that the slope has changed to a positive one in 1995. The cumulative curves in Figure 10 change from convexity to concavity in 1995. Due to the extraordinary predictive power and reliability of all empirical relationships obtained in this study, we ignore the assumption that there is no linear link between unemployment, inflation, and labour force and that the deviation started in 1995 reflects unpredictable and spontaneous character of all involved variables.

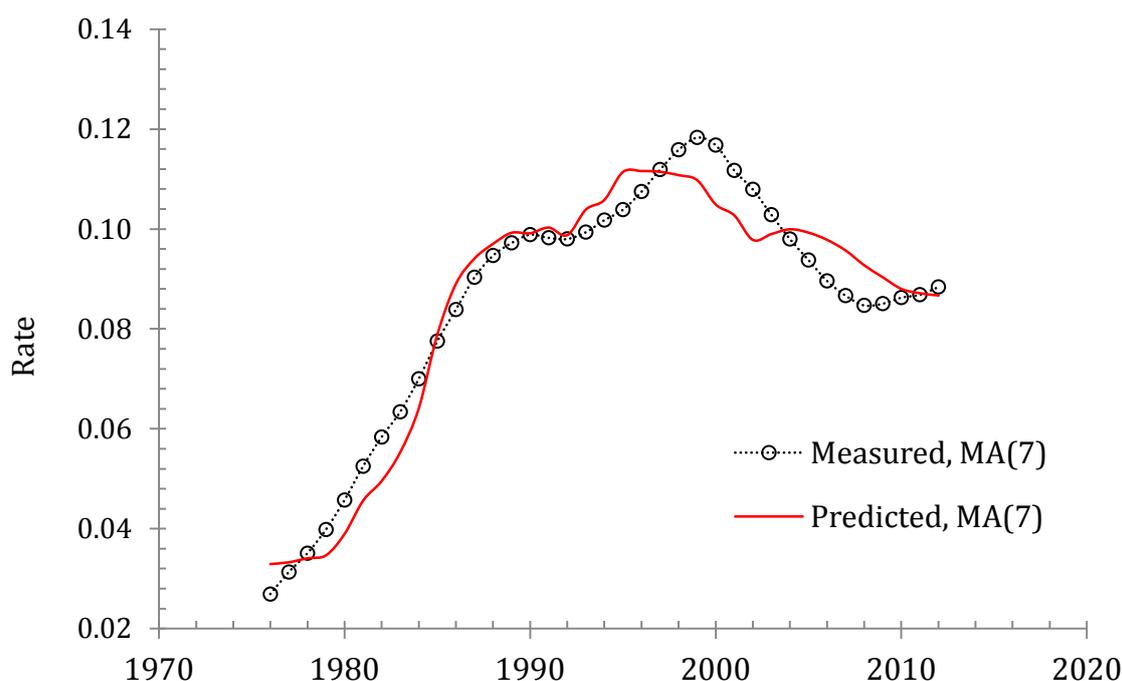

Figure 11. The predicted and measured rate of unemployment both smoothed with MA(7).

Figure 11 displays the annual measured and predicted rate of unemployment smoothed with MA(7). These curves practically coincide. There is also a significant improvement in the predictive power of relationship (13). The overall agreement is also improved as reflected by a higher $R^2$=0.95 and lower standard error $\sigma$=0.005. Moving average and other smoothing techniques are good for noise suppression but may bias OLS regression analysis as associated with data overlapping when synchronized time series are regressed. Hence, accurate measurements of labour force are mandatory for obtaining higher correlation between observed and predicted time series. In this regard, more accurate measurements is a substitute of moving average.

Finally, we consider the rate of inflation, unemployment, and the change in labour force altogether. For France, the generalized relationship is obtained as a sum of (10) and (13), which results, with some marginal tuning of all coefficients in order to reduce the standard error of the model, in the following equation for the GDP deflator:

$$\pi(t) = 2.69l(t-5) - u(t-5) + 0.108; \quad 1971 \leq t \leq 1995$$
$$\pi(t) = 6.40l(t-5) - u(t-5) + 0.059; \quad t \geq 1996 \quad (14)$$

For the OECD CPI:

$$\pi(t) = 3.0l(t-5) - u(t-5) + 0.108; \quad 1971 \leq t \leq 1995$$
$$\pi(t) = 5.0l(t-5) - u(t-5) + 0.067; \quad t \geq 1996 \quad (15)$$

where we model inflation since it lags by 5 years behind the change in labour force and unemployment. Formally, one can re-write both relationships for $u(t)$. Notice that the change in the slopes and intercepts are much smaller than in individual relationships. The structural break is less prominent and thus its estimate is less reliable.

The annual and cumulative curves for both cases are presented in Figure 12. Linear regression of the observed inflation against that predicted according to (14) and (15) is characterized by outstanding for annual curves statistical properties: $R^2$=0.87 and RMSFE=0.015 $y^{-1}$, and $R^2$=0.83 and RMSFE=0.017 $y^{-1}$, respectively. For the cumulative curves, both $R^2$ are larger than 0.99 and RMSFE~0.025 $y^{-1}$, i.e. by 20% smaller than the naive ones (see Table 4). These estimates were obtained for the period between 1972 and 2012 with a five-year lag. These RMSFEs are the best obtained for France at a five year horizon so far. They explain the rate of price inflation to the extent beyond which measurement uncertainty should play the key role. Practically, there is no room for any further improvements in $R^2$ given the accuracy of the current prediction.

## 5. Conclusion

We have successfully modelled unemployment and inflation in France. Their sensitivity to the change in labour force requires very accurate measurements for any quantitative modelling to be reliable. Unfortunately, the OECD labour force time series does not meet this requirement and poor statistical results are obtained for annual readings. The best prediction is obtained with the moving average technique applied to the change in labour force. For the period between 1970 and 2012, linear regression analysis provides $R^2$ as high as 0.8 to 0.9 for the rate of unemployment and GDP deflator. The RMSFE for the best CPI model is 0.015 $y^{-1}$ and 0.010 $y^{-1}$ for the GDP deflator, both at a four year horizon. For the period after 1994, the best RMSFE=0.005 $y^{-1}$ for both measures of inflation. In 1994, our models have structural breaks found by the OLS fit. For the VECM representation, the standard error for the GDP deflator is as low as 0.010 $y^{-1}$ at a four year horizon and 0.005 $y^{-1}$ for a two year horizon. The whole period and 0.004 $y^{-1}$ for the period after 1994. All in all, we have obtained a very accurate description of unemployment and inflation in France during the past 40 years.

Having discussed the technically solvable problems associated with the uncertainty in the labour force measurements, we start tackling the problem associated with the divergence of the observed and predicted curves starting around 1995. An understanding of this discrepancy is a challenge for our concept. Potentially, these curves diverge due to the new monetary policy introduced by the Banque de France. We may claim that the policy of constrained money supply, if applied, could artificially disturb relationships (9), (10), and (13). We had to introduce a structural break and to estimate new coefficients after 1995 for unemployment and after 1994 for inflation, respectively. These coefficients are less reliable because the relevant time series are short and vary in narrow dynamic ranges, but they are definitely different from those before the breaks. One could conclude that Banque de France has created some new links between the unemployment, inflation, and labour force, shifting coefficients in the original long term equilibrium relations.

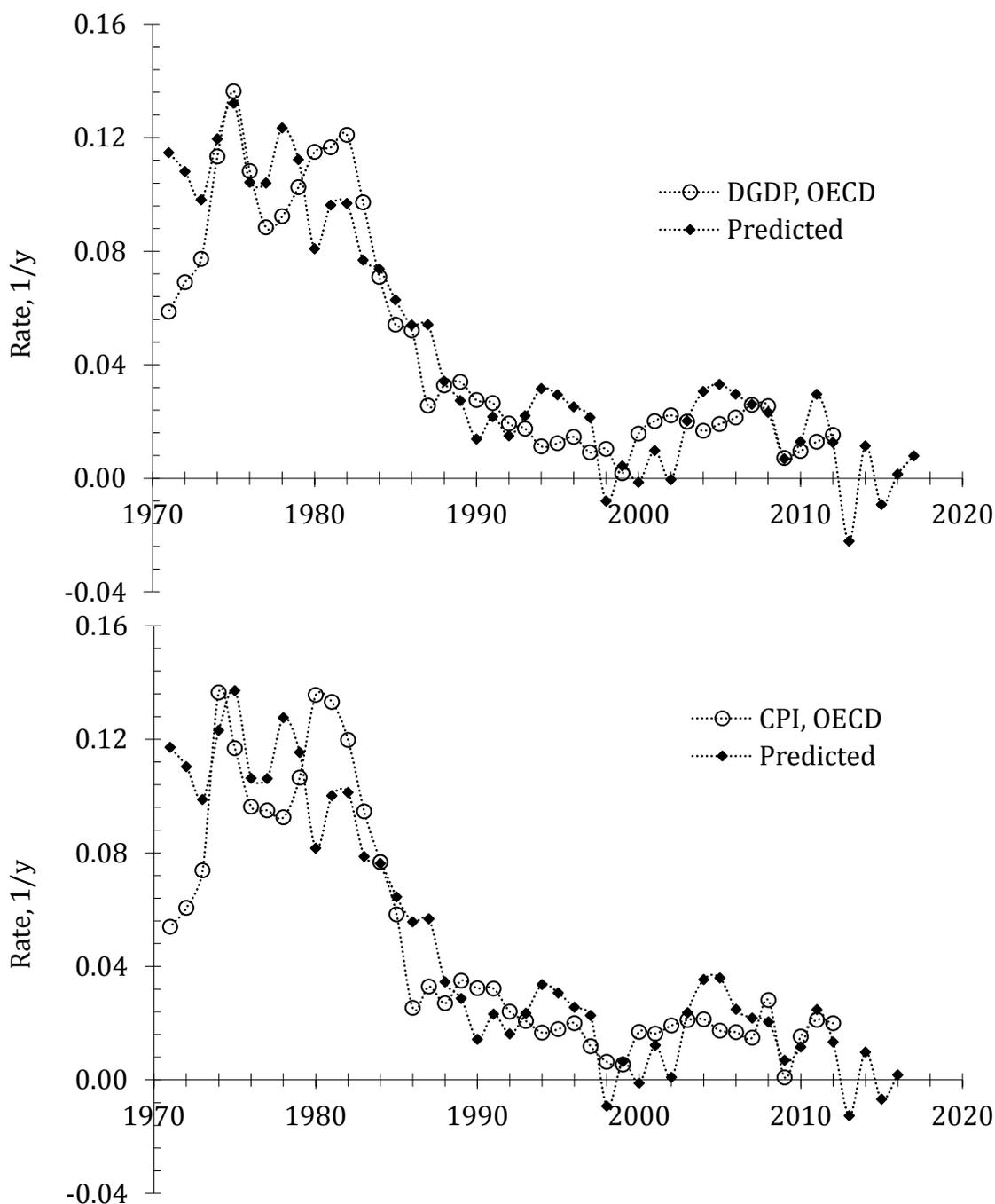

Figure 12. Comparison of the observed and predicted inflation in France - annual and cumulative inflation since 1972. The predicted inflation is a linear function of the labour force change and unemployment.

We think that the true money supply in excess of that related to real GDP growth should be completely controlled by the demand related to the growing labour force. This excessive money supply is accommodated in developed economies through employment growth, which then causes price inflation. The latter serves as a mechanism effectively returning the normalized personal income distribution to its original shape (Kitov and Kitov, 2013). The relative amount of money that the economy needs to accommodate through increasing employment, as a reaction on independently growing labour force, is constant through time but varies among developed countries. This amount has to be supplied to the economy by central bank.

The ESCB limits money supply to achieve price stability. For France, the growth in labour force was so intensive after 1995 that it requires a much larger money supply for creation of an appropriate number of new jobs. The 2% artificial constraint on inflation, and thus on the money supply, disturbs relationships (10) and (13). Due to lack of money in the French economy, the actual (and mainly exogenous) growth in labour force was only partially accommodated by 2% inflation. The lack of inflation resulted in increasing employment. In other words, instead of 2% unemployment, as one should expect according to the relationship before 1995, France had 9% unemployment. Those people who entered the labour force in France in excess of that allowed by the target inflation rate had no choice except to join unemployment in order to compensate the natural 7% rate of inflation, which was suppressed to 2%.

The lags and amplification factors (sensitivities) found for unemployment and inflation in France are quite different from those obtained for the USA and Austria (Kitov and Kitov, 2010). The latter country is characterized by the absence of time lags and low sensitivities. In the USA, inflation lags by two and unemployment by five years behind the change in labour force, with sensitivities much lower than those in France. Apparently, the variety of lags is the source of problems for the Phillips curve concept.

The causal link between inflation, unemployment, and labour force gives a unique opportunity to foresee future at extra long time horizons. The accuracy of such long-term unemployment and inflation forecasts is proportional to the accuracy of labour force projections. For example, central banks can use labour force projections as a proxy to "inflation expectation" in their NKPCs. Figures 8 and 12 imply that France will be enjoying a period of low inflation rate in the near future. Monetary policy of the ECB is also an important factor for these forecasts because of its influence on the partition of the labour force growth between inflation and unemployment. Moreover, this is the responsibility of the ECB and Banque de France to decide on the partition.